\documentclass[journal,twoside,web]{ieeecolor}
\usepackage{generic}
\usepackage{cite}
\usepackage{amsmath,amssymb,amsfonts}
\usepackage{algorithmic}
\usepackage{graphicx}
\usepackage{algorithm,algorithmic}
\usepackage{hyperref}
\hypersetup{
    colorlinks=true,
    linkcolor=blue,
    citecolor=blue,
    urlcolor=blue
}
\usepackage{textcomp}

\usepackage{textcomp}
\usepackage{booktabs}
\usepackage{color}
\usepackage{multirow}
\usepackage{threeparttable}
\usepackage{pifont}
\usepackage{subfig}

\def\BibTeX{{\rm B\kern-.05em{\sc i\kern-.025em b}\kern-.08em
    T\kern-.1667em\lower.7ex\hbox{E}\kern-.125emX}}
\begin{document}
\title{A Real-time Scale-robust Network for Glottis Segmentation in Nasal Transnasal Intubation}
\author{Yang Zhou, Chaoyong Zhang, Ruoyi Hao, Huilin Pan, Yang Zhang and Hongliang Ren, \IEEEmembership{Senior Member, IEEE}
\thanks{Manuscript received xx xx, 2025; revised xx xx, xxxx. 
The work was supported in part by the National Key R\&D Program of China under Grant 2018YFB1307700 (also with subprogram 2018YFB1307703) from the Ministry of Science and Technology (MOST) of China, Hong Kong Research Grants Council (RGC) Collaborative Research Fund (CRF C4026-21GF and CRF C4063-18G), and General Research Fund (GRF \#14211420); Shun Hing Institute of Advanced Engineering (BME-p1-21/8115064) at the CUHK; Shenzhen-Hong Kong-Macau Technology Research Program (Type C) Grant 202108233000303. (Hongliang Ren and Y. Zhang are co-corresponding authors.) \{Corresponding to: Hongliang Ren. hlren@ieee.org; Yang Zhang. yzhangcst@hbut.edu.cn}
\thanks{Y. Zhou and C. Zhang are with the School of Mechanical Science \& Engineering Huazhong University of Science and Technology, Wuhan 430074, China; H. Pan is with the School of
Artificial Intelligence and Automation Huazhong University of Science and Technology(e-mail: yzhoucst@hust.edu.cn; zcyhust@hust.edu.cn; hlp@hust.edu.cn)}
\thanks{Y. Zhang is with the School of Mechanical Engineering, Hubei University of Technology, Wuhan 430068, China; and the National Key Laboratory for Novel Software Technology, Department of Computer Science and Technology, Nanjing University, Nanjing 210023, China. (e-mail: yzhangcst@hbut.edu.cn)}
\thanks{H. Ren and R. Hao are with the Department of Electronic Engineering, The Chinese University of Hong Kong, Hong Kong, China; and with the Shun Hing Institute of Advanced Engineering, The Chinese University of Hong Kong, Hong Kong 999077, China; and with the Department of Biomedical Engineering, National University of Singapore (NUS), Singapore 117575, Singapore, and NUS (Suzhou) Research Institute, Suzhou 215123, China. (e-mail: hlren@ieee.org; 1155167067@link.cuhk.edu.hk)}}
\maketitle

\begin{abstract}
Nasotracheal intubation (NTI) is a critical clinical procedure for establishing and maintaining patient airway patency. 
Machine-assisted NTI has emerged as a pivotal approach for optimizing procedural efficiency and minimizing manual intervention.
However, visual detection algorithms employed for NTI navigation encounter significant challenges, including complex anatomical environments and suboptimal illumination conditions surrounding the glottis.
Additionally, the glottis presents considerable scale variability throughout the procedure, initially appearing as a small, difficult-to-capture structure before expanding to occupy nearly the entire field of view. 
Moreover, traditional visual detection methods often have high computational costs, making real-time, high-precision detection on portable devices challenging. 
To enhance NTI efficacy and address these challenges, this paper proposes a novel glottis segmentation framework optimized for vision-assisted NTI applications.
First, we designed a lightweight, multi-receptive field feature extraction module to reduce intra-class differences, achieving robustness to scale variations of the glottis.
This module was then stacked to form the backbone and neck of our network. 
Subsequently, we developed an advanced label assignment method and redefined the number of samples to further reduce intra-class differences and enhance accuracy in the complex NTI environment. 
Experiments on three distinct datasets demonstrate that our network surpasses state-of-the-art algorithms, achieving a segmentation mDice of 92.9\% with a compact model size of 19 MB and an inference speed exceeding 170 frames per second.
Our code and datasets are available
at https://github.com/HBUT-CV/GlottisNet. 
\end{abstract}

\begin{IEEEkeywords}
Convolutional neural network, Medical image segmentation, Endoscopic navigation, Nasal transnasal intubation
\end{IEEEkeywords}

\section{Introduction}
\label{sec:introduction}
\IEEEPARstart{N}{asotracheal} intubation (NTI) represents a critical clinical procedure for establishing and maintaining patient airway patency.
Unlike routine orotracheal intubation, NTI is specifically indicated for patients presenting with restricted mouth opening, cervical spine instability, or those undergoing maxillofacial surgeries where the surgical field precludes oral access, as outlined in the \textit{2022 ASA Practice Guidelines for Management of the Difficult Airway}~\cite{ASA}.
However, this procedure demands substantial clinical expertise and fine motor control, often resulting in a relatively low success rate for manual NTI~\cite{JBHI1, TMI2}.
Repeated intubation attempts or procedural failure can lead to severe life-threatening complications, including hypoxemia, aspiration, and irreversible airway trauma.
Furthermore, in challenging environments such as wilderness rescue scenarios, NTI procedures are frequently performed by emergency medical technicians with limited specialized training rather than by expert clinicians.
In response to these challenges, machine-assisted visual navigation has emerged as a pivotal approach to standardize procedural outcomes, reduce dependency on operator experience, and optimize the overall efficiency of airway management.

While recent advances in robotics and computer vision have facilitated automated navigation, existing solutions still face significant limitations. Current state-of-the-art (SOTA) visual detection algorithms primarily rely on the direct application of general-purpose object detectors. For instance, lightweight single-stage detectors often sacrifice segmentation precision for speed, failing to precisely delineate the fine boundaries of the glottis under extreme scale variations. Conversely, high-precision two-stage frameworks incur high computational costs, rendering them unsuitable for deployment on portable medical devices. 
Thus, there remains a critical need for a specialized framework capable of simultaneously achieving high-precision glottis segmentation and real-time inference on edge devices.

Drawing inspiration from RTMDet~\cite{lyu2022rtmdet}, we propose a lightweight glottis detection architecture that optimizes visual navigation in robotic intubation procedures.
The proposed architecture enhances detection efficiency while significantly reducing computational overhead.
Such improvements enable model deployment on edge computing devices, thereby facilitating seamless integration between visual detection hardware and robotic control systems.
Nevertheless, the glottis opening occupies merely a minute fraction of the bounding box area; consequently, navigating based solely on the bounding box center may result in misalignment with the actual glottic position, compromising navigational precision.

Instance segmentation enables precise pixel-level localization of the glottis, thereby providing high-fidelity navigational guidance.
Critically, incorporating additional instance segmentation branches introduces substantial computational cost and complicates the model deployment process.
Moreover, throughout the NTI procedure, the glottis exhibits considerable scale variations within the endoscopic view.
Additionally, the periglottic environment is characterized by complex morphological features and inadequate illumination.
Under these challenging conditions, conventional instance segmentation networks frequently demonstrate significant performance degradation when segmenting targets with drastic scale disparities.

To address these challenges, we introduce the \textbf{\textit{Light}}weight \textbf{\textit{S}}cale \textbf{\textit{R}}obust \textbf{\textit{M}}odule (LightSRM). 
This module leverages multi-scale receptive fields for feature extraction, effectively minimizing intra-class variations while maintaining computational efficiency and robust performance across varying glottis scales.
Afterward, we integrated LightSRM into the backbone and incorporated it within the path aggregation network (PAN) to enhance feature fusion capabilities.
To enable the network to perform segmentation, we implemented an additional prediction branch for mask generation.
Finally, to mitigate the effects of low illumination conditions and complex anatomical scenes during intubation, we developed an optimized label assignment method with carefully calibrated weights and balanced positive-negative sample ratios, thereby further reducing intra-class variations and enhancing overall system performance.

The proposed approach was comprehensively evaluated using three distinct datasets: a publicly available dataset, a custom-developed phantom image dataset, and a clinical dataset, with performance benchmarked against state-of-the-art methods.
The proposed model achieves superior performance in both segmentation and detection tasks while maintaining a compact model size of 19 MB.
With inference speeds exceeding 170 frames per second (FPS) and consistently high detection accuracy, the proposed model satisfies the stringent requirements for NTI deployment in resource-limited emergency scenarios.
The key contributions of this work are summarized as follows:

\begin{enumerate}[]
    \item Develope a lightweight real-time segmentation framework that enables rapid detection of multi-scale glottal structures in complex environments, providing precise visual navigation for robot-assisted NTI.
    \item Propose a novel multi-receptive field feature extraction module (LightSRM) that effectively mitigates intra-class variations and maintains robust performance across diverse glottic scale variations.
    \item Through meticulously designing the sample quantity and implementing a novel label assignment method, we minimize environmental impact and further reduce intra-class differences, thereby improving accuracy.
    \item Results on three comprehensive datasets show that our method achieves state-of-the-art detection accuracy, while maintaining a compact model size (19 MB) and superior inference speed ($>$170 FPS).
\end{enumerate}

The remainder of this paper is organized as follows:
Section~\ref{related_works} introduces the related work on detection and segmentation. 
Section~\ref{method} details our proposed methodology.
Section~\ref{experiments} presents extensive ablation studies evaluating the contribution of individual components within our framework, along with detailed descriptions of our experimental setup and custom-developed datasets.
Finally, Section~\ref{conclusion} concludes this work.

\begin{figure*}[!t]
\centering
\centerline{\includegraphics[width=6.9in]{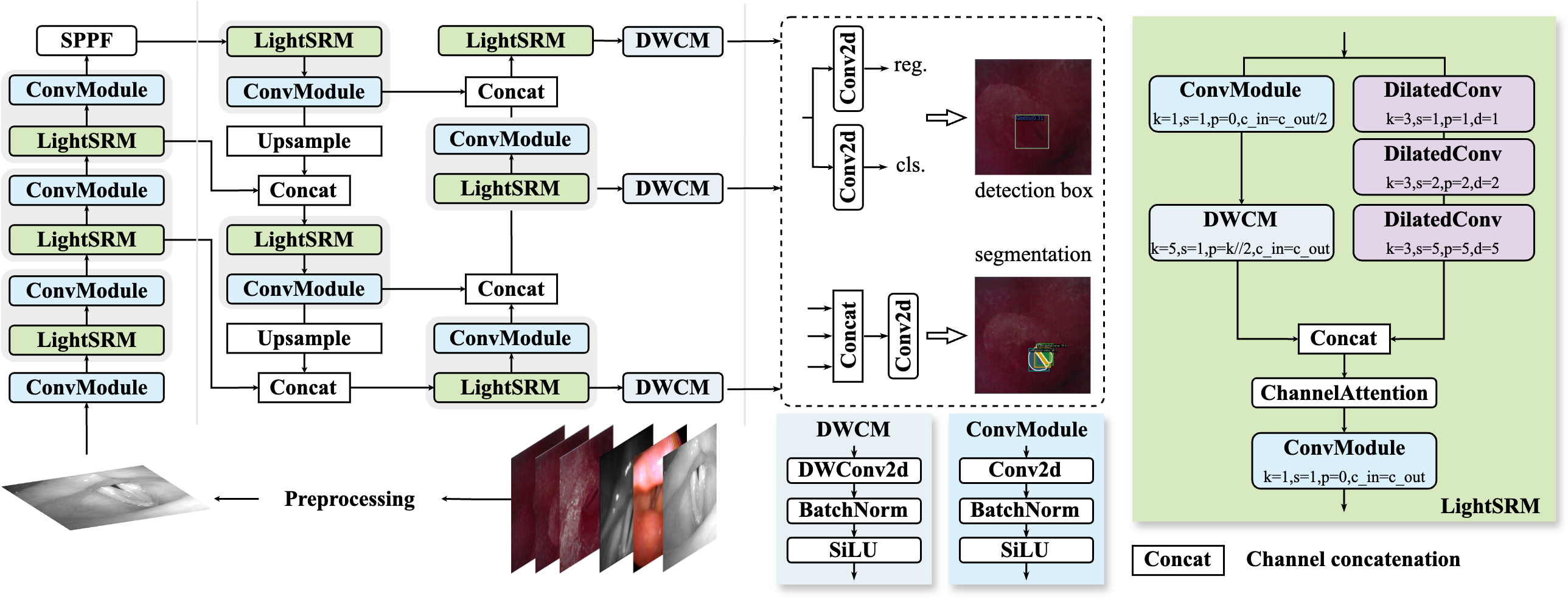}}
\caption{The overview of the proposed framework. Our framework includes several parts: the convolution module (ConvModule), the lightweight scale robust module  (LightSRM), the depthwise convolution module (DWCM), and other modules such as concatenation (Concat) and spatial pyramid pooling fast (SPPF) layers. 
LightSRM consists of a main path and a shortcut. The main path employs cascaded dilated convolutions to achieve a larger receptive field. The shortcut path uses $5\times5$ depthwise separable convolutions to reduce computation. The feature maps from both paths are concatenated and passed through a channel attention module to enhance inter-channel interaction. Finally, a $1\times1$ convolution is used to adjust the number of channels.}
\label{overall}
\end{figure*}

\section{Related Works} 
\label{related_works}
\subsection{Glottis Detection and Segmentation}
In vision-assisted intubation and computer-aided diagnosis, automated detection and segmentation of the glottis have become focal points of research.
Early studies~\cite{G2, Lai, G3} employed conventional image processing techniques, but these exhibited limited segmentation accuracy and inadequate robustness to environmental variations such as lighting changes and motion blur.
With the advancement of deep learning, research has shifted towards convolutional neural network (CNN)-based methodologies, particularly within the domain of robot-assisted intubation systems.

For instance, Wang et al.~\cite{I1} developed a robot-assisted system wherein a robotic arm holds a fiber laryngoscope while an operator remotely guides the insertion trajectory. Although this approach facilitates procedures in hazardous environments, its effectiveness remains heavily dependent on operator expertise due to the lack of closed-loop visual feedback. To achieve automatic navigation, Deng et al.~\cite{I2} employed a CNN to generate heading targets for NTI. However, their approach achieved a limited accuracy of 79.4\% with an inference speed of merely 13 FPS, failing to meet the real-time performance benchmarks essential for clinical implementation.

Regarding specific glottis segmentation algorithms, recent studies have applied CNN variants to similar tasks, such as swallowing evaluation. Researchers have utilized attention-based frameworks to segment the glottal valves and the glottic slit~\cite{FEES1, FEES2}. Given the profound impact of U-Net~\cite{U-Net} on medical image segmentation, numerous studies have adopted U-Net-based architectures for glottis segmentation. However, standard medical segmentation models like U-Net, U-Net++~\cite{unet++}, and U-Net 3+~\cite{unet3+} often suffer from large model sizes and slow inference speeds, rendering them unsuitable for deployment on portable edge devices in field scenarios.

To support the development of data-driven algorithms, \cite{Dataset1} and \cite{Dataset2} established comprehensive glottal segmentation datasets utilizing real clinical data. These datasets provide a crucial foundation for subsequent investigations. Despite these advances, current visual assistance intubation algorithms have yet to achieve an optimal balance among real-time processing capabilities, scale robustness, and high-precision performance.

\subsection{Object Detection and Segmentation}
Two-stage frameworks like Mask R-CNN~\cite{maskrcnn} and its variants~\cite{Faster-RCNN, Cascade_RCNN} established strong baselines for instance segmentation but incur high computational costs, impeding real-time deployment. Subsequent methods like CondInst~\cite{CondInst} and BoxInst~\cite{BoxInst} improved efficiency by using dynamic kernels or box-only supervision, yet they still struggle to balance speed with the high-precision boundary delineation required for glottis segmentation.

While Mask R-CNN-based two-stage frameworks exhibit high model complexity that impedes real-time detection, single-stage detectors such as RTMDet~\cite{lyu2022rtmdet} and YOLACT~\cite{YOLACT}, as well as the YOLO series~\cite{yolov8_u, wang2024yolov9, yolov10_u, yolo11_u}, inherently compromise precise boundary delineation. Although these models demonstrate superior object detection performance, their suboptimal pixel-level capabilities restrict their utility in applications demanding high-precision segmentation. Furthermore, despite accelerating network convergence, architectures including DETR~\cite{DETR} and its variants~\cite{Conditional_DETR, DAB_DETR, Deformable_DETR}, alongside approaches like GFL~\cite{GFL}, SOLOv2~\cite{SOLOv2}, SparseInst~\cite{SparseInst}, and ATSS~\cite{ATSS}, exhibit comparable structural limitations in highly accurate instance segmentation.

Although visual foundation models including SAM~\cite{SAM}, MedSAM~\cite{MedSAM}, and FastSAM~\cite{FastSAM}, along with recent adaptations~\cite{RSPrompter, MedSAM2, MedSAM_Adapter}, significantly advance medical image segmentation, their massive parameter scales incur prohibitive computational overhead. This severely limits low-latency inference in resource-constrained clinical environments. Consequently, developing a lightweight, efficient, and accurate segmentation model remains critical to this study.

\section{Algorithm}
\label{method}
\subsection{Overall Framework}
We propose a lightweight, scale-robust framework for real-time glottis segmentation (GlottisNet), with its architectural overview illustrated in Fig.~\ref{overall}.
Unlike conventional architectures, our network introduces a unified design paradigm wherein both the backbone and neck components utilize identical LightSRM modules. The comprehensive architecture of LightSRM is detailed in the subsequent section. This modular consistency across backbone and neck components substantially optimizes network design efficiency. 

The neck adopts a PAN-inspired design and leverages LightSRM modules to implement bidirectional feature propagation pathways (both top-down and bottom-up), enhancing multi-scale feature fusion.
The architecture employs depthwise convolution modules to normalize the feature map dimensions across channels, ensuring consistent dimensionality across all three output layers for balanced feature representation. 

To maximize computational efficiency and minimize model complexity, we implemented a streamlined detection head design.
Benefiting from the superior design of LightSRM, the backbone and neck exhibit robust feature extraction and fusion capabilities. 
This architectural efficiency eliminates the need for additional feature processing in the head, employing only $1 \times 1$ convolutions for feature map dimensionality reduction and direct task-specific predictions (classification, regression, and mask generation).
This streamlined architecture substantially reduces computational overhead while maintaining detection efficacy.

\subsection{Basic LightSRM Block}
\begin{figure}[!t]
    \centerline{\includegraphics[width=3.4in]{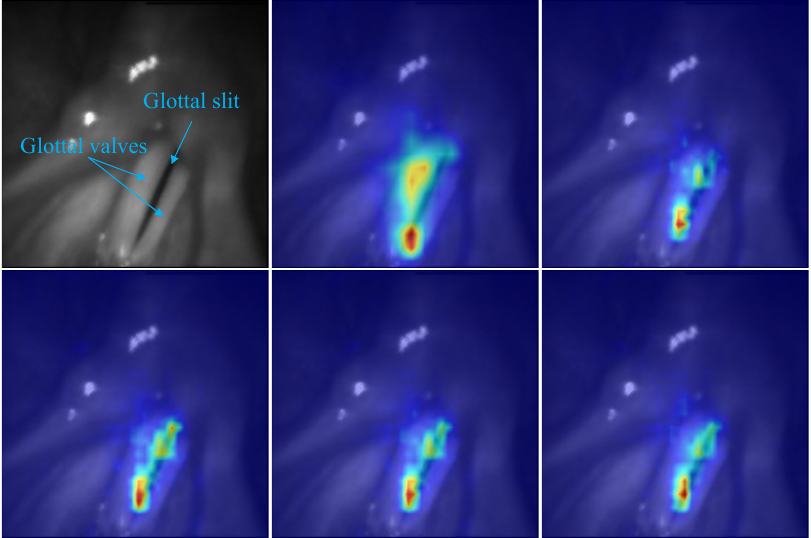}}
    \caption{Grad-CAM activation heatmaps illustrating feature evolution within the first LightSRM at the bottom of the FPN. The first row compares the annotated raw input, the input heatmap of the LightSRM, and the output of the depthwise convolution. The second row shows the heatmap after the DilatedConv, the concatenation operation, and the attention module. LightSRM combines a large receptive field with precise positional information, focusing on the glottic gap, thereby improving detection accuracy.}
    \label{fig4}
\end{figure}

To reduce intra-class variance and achieve robustness to glottis scale variations, we propose LightSRM, a novel architectural module illustrated in Fig.~\ref{overall}.

The main path employs a cascade of three dilated convolutions to generate an expanded receptive field, and the receptive field size is calculated as follows:
\begin{equation}
    R_{j+1}=R_{j}+(k_{j+1}-1)\ast\prod_{i+1}^{j}s_i.
\end{equation}
Here, $R_{j+1}$ denotes the receptive field size of the current feature map, $k_{j+1}$ represents the kernel size of the current convolutional layer, and $s$ is the product of the strides of the previous convolution layers~\cite{Dliate}.
Meanwhile, to avoid receptive field holes affecting information acquisition at various distances, we systematically determine the kernel sizes for cascaded dilated convolutions using:
\begin{equation}
    D_{i}=max[D_{i+1}-2r_i, 2r_i-D_{i+1}, r_i],
\end{equation}
where $r_i$ denotes the dilation rate at layer $i$, and $D_{i}$ represents the maximum inter-element distance within that layer. Through theoretical analysis, we derive optimal dilation rates $[1, 2, 5]$, which are implemented in the LightSRM architecture.
In this configuration, the receptive field size in the main path of LightSRM is $17 \times 17$. Under equivalent parameter constraints, if the main path is replaced with three standard $3 \times 3$ convolutions, the receptive field is limited to $7 \times 7$.

The shortcut path incorporates a $5 \times 5$ depthwise separable convolution for efficient feature extraction, optimizing computational complexity while preserving low-dimensional feature propagation. Both the main path and shortcut of LightSRM maintain the original dimensions of the feature maps.

Subsequently, the outputs from both paths are concatenated along the channel dimension, thereby enhancing the complexity of the feature maps. A channel attention module is then applied to the concatenated output, facilitating enhanced inter-channel interactions. Finally, a $1 \times 1$ convolution layer modulates the channel dimensionality, ensuring consistency between input and output features. By integrating multiple receptive fields, this module effectively reduces inter-class variations while achieving substantial improvements in accuracy at minimal computational cost. 

We use Grad-CAM\cite{CAM} to generate activation heatmaps for the glottic slit based on the feature maps of the first LightSRM at the bottom layer of the neck in Fig.~\ref{fig4}.
The first row presents three sequential visualizations: the annotated input image, the input heatmap of LightSRM, and the output following depthwise convolution. The second row illustrates the progressive feature transformations through dilated convolution (DilatedConv), the concatenation operation, and the attention module. While the pathway utilizing large convolution kernels effectively directs attention toward the glottic gap, it fails to achieve comprehensive glottis coverage.

The implementation of dilated convolutions enables enhanced emphasis on the glottic gap, demonstrating the advantages of an expanded receptive field.
Following feature map concatenation and channel attention processing, the attention of the network encompasses the entire glottic gap, with a specific focus on the intubation site. 
The visualization in Fig.~\ref{fig4} demonstrates the effectiveness of the LightSRM design in facilitating critical feature learning.

\subsection{Scale robust feature pyramid network}

\begin{figure}[!t]
    \centerline{\includegraphics[width=3.4in]{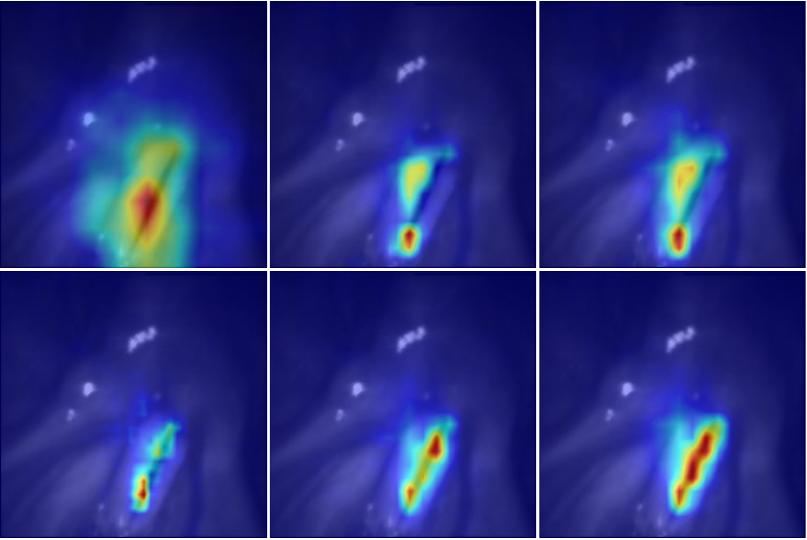}}
    \caption{Grad-CAM activation heatmaps illustrating feature evolution within the overall network. The above and below sequences represent the heatmaps in the bottom-up and top-down pathways, respectively. Through the SRFPN feature fusion mechanism, the attention field undergoes progressive spatial refinement, ultimately converging on the anatomically relevant glottal aperture region.}
    \label{fig5}
\end{figure}

To mitigate the accuracy fluctuations caused by scale variations of the glottis during NTI, we enhance feature fusion in the neck by introducing the \textbf{\textit{S}}cale \textbf{\textit{R}}obust \textbf{\textit{F}}eature \textbf{\textit{P}}yramid \textbf{\textit{N}}etwork (SRFPN). Inspired by RTMDet~\cite{lyu2022rtmdet}, we augment the Feature Pyramid Network (FPN) with a bottom-up pathway analogous to the Path Aggregation Network (PAN). This bidirectional architecture facilitates the integration of fine-grained lower-level features with semantic-rich higher-level representations, optimizing information flow.

To optimize architectural efficiency and maintain balanced parameter distribution between the backbone and neck, we incorporate LightSRM within the neck. For computational efficiency, we implement three $3 \times 3$ depthwise separable convolutions at the SRFPN output stage for channel dimensionality adjustment. This design substantially reduces computational complexity while maintaining uniform channel dimensionality across all three SRFPN outputs.

To more intuitively demonstrate the superiority of our neck design, we present activation heatmaps visualizations following each LightSRM along both top-down and bottom-up pathways.
As illustrated in Fig.~\ref{fig5}, the backbone successfully captures the coarse spatial location and morphological characteristics of the glottis.
The initial LightSRM refines feature representations, enhancing network attention specifically toward the glottic region.
Subsequent LightSRMs progressively enhance feature discrimination in both inferior and superior regions of the glottic aperture. 
The final LightSRM achieves precise localization of the glottic gap while effectively suppressing irrelevant peripheral features.
The final network output demonstrates comprehensive learning of glottic characteristics, exhibiting uniform attention distribution across the glottic aperture.

\subsection{Label Assignment}
We introduce an adaptive label assignment method that employs dynamic sample matching strategies to optimize the cost-based selection of positive and negative samples.
This method implements fine-grained sample quality discrimination mechanisms, which effectively minimize intra-class differences while preventing accuracy losses and non-convergence issues arising from scale variations.

Following ATSS~\cite{ATSS}, we compute the label assignment cost by incorporating three components: the classification cost ($C_{cls}$), the intersection over union (IoU) cost ($C_{IoU}$), and the center priority cost ($C_{cen\_pre}$).
The cost matrix can be formulated as follows:
\begin{equation}
    C=\lambda_{1} C_{IoU} + \lambda_{2} C_{cls} + \lambda_{3} C_{cen\_pre},
\end{equation}
where $\lambda$ represents the weights of each cost, we conducted detailed ablation experiments in Table \ref{table2}, ultimately setting $\lambda_{1}=3$, $\lambda_{2}=1$, and $\lambda_{3}=3$.

Initially, we utilize center priority to establish the sampling range for high-quality positive samples. 
The center prior cost is calculated based on the Euclidean distance between the geometric center points of the predicted and ground truth (GT) bounding boxes.
During NTI, the flexible endoscope requires continuous maneuvering, including rotation and bending, to navigate the tortuous anatomical structures of the nasal cavity and pharynx. Consequently, the glottis appears at unpredictable positions and orientations within the captured 2D endoscopic view.
We employ the Euclidean distance as a rotation-invariant metric to mitigate the impact of unpredictable positional shifts and in-plane rotations of the glottis caused by endoscopic maneuvering. We refer to these geometric changes as glottic variations. The center priority cost is computed as follows:
\begin{equation}
C_{\text{cen\_pre}} = 10^{ \| \text{center}_{pre_{box}} - \text{center}_{\text{GT}} \|_2 - \xi},
\end{equation}
where $\text{center}_{pre_{box}}$ and $\text{center}_{\text{GT}}$ represent the centroid coordinates of the prediction and the GT, respectively. 
The exponential function with base 10 acts as a soft gating mechanism. It imposes a steep penalty gradient on samples that significantly deviate from the center and effectively suppresses low-quality candidates. The hyperparameter $\xi$ (set to 3) serves as a tolerance radius to ensure that predictions within this spatial margin are assigned lower costs to encourage convergence.

We also illustrate the changes in attention brought about by the center prior in Fig.~\ref{fig5}. 
The attention initially concentrates on the glottic aperture, progressively expanding its focal distribution through deeper network layers until achieving comprehensive coverage of the target region.
Regarding classification cost calculation, soft labels are defined based on the IoU between the predicted bounding box ($pre_{box}$) and the ground truth (GT).
An IoU threshold of 0.5 determines the binary classification: values exceeding this threshold are designated as positive, while others are classified as negative.
The calculation of the soft label $y_{soft}$ is as follows:
\begin{equation}
y_{soft}= \begin{cases} 
1, & \text{if } IoU(pre_{box}, \text{GT})>0.5 \\
0, & \text{otherwise.}
\end{cases}
\end{equation} 
The classification cost utilizes the generalized focal loss, aggregating contributions from both positive and negative samples:
\begin{equation}
C_{cls}=-y_{soft}(1-p)^{\gamma}\log(p) - (1-y_{soft})p^{\gamma}\log(1-p),
\end{equation}
where $p$ represents the predicted classification probability and the latter $\gamma$ is the adjustment factor for easy-to-classify samples, generally set to 2.

To optimize computational efficiency while maximizing sample quality discrimination, the IoU cost utilizes the previously computed IoU values from the classification stage.
Additionally, to make the IoU cost smoother and further increase the distinction between samples, we apply the log function to amplify the IoU values. The calculation method is as follows:
\begin{equation}
C_{IoU}=-log(IoU + \epsilon),
\end{equation} 
where $\epsilon$ is a small positive constant that ensures numerical stability by preventing logarithm evaluation at zero.

\section{Experiments}
\label{experiments}
\subsection{Hardware System} 
\begin{figure}[!t]
\centerline{\includegraphics[width=\columnwidth]{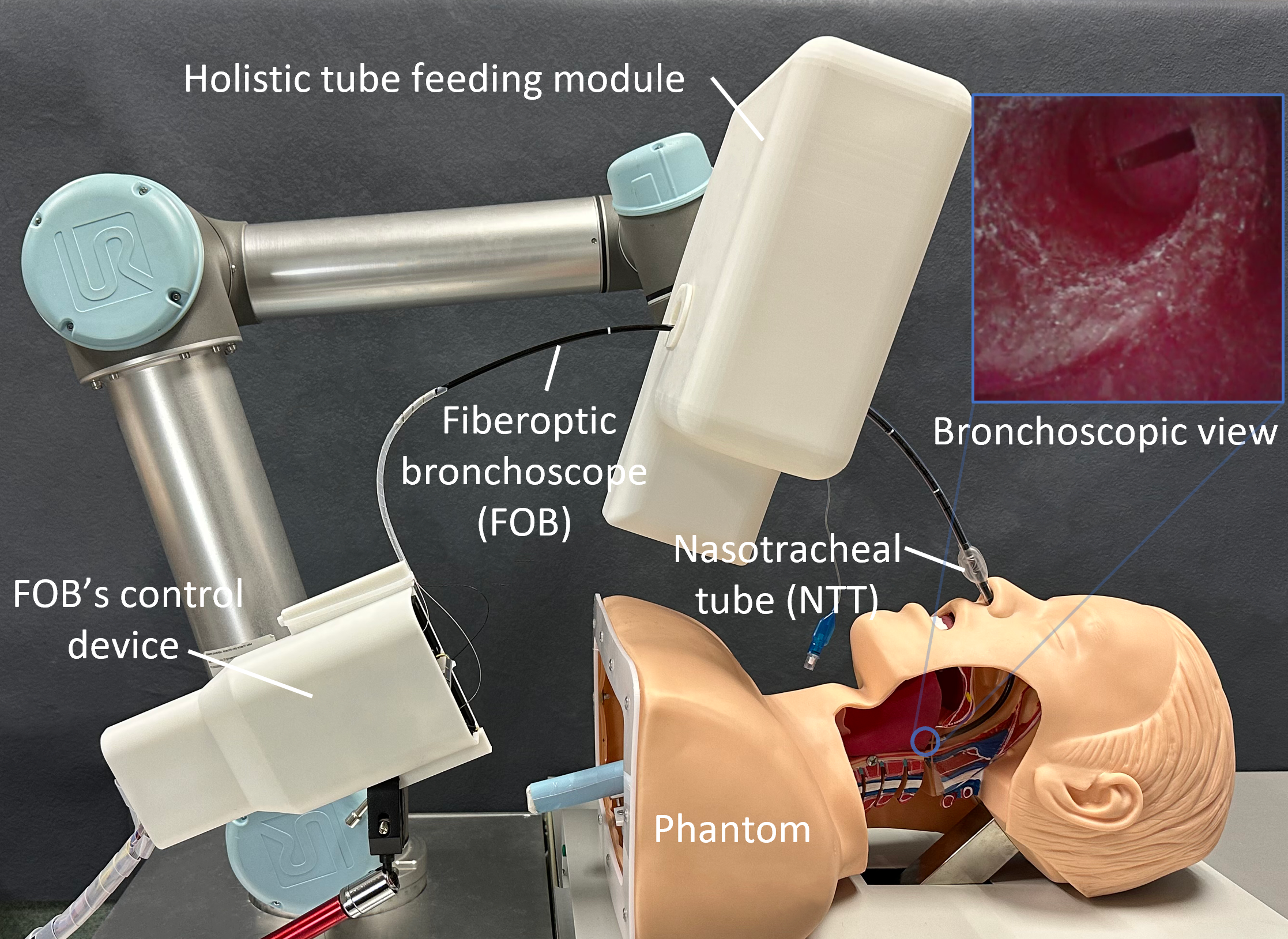}}
\caption{Nasotracheal intubation robot system. The system determines the insertion direction by controlling the bending angle of the FOB and uses a robotic arm to remotely guide the FOB for intubation.}
\label{hardware}
\end{figure}

To verify the performance of the proposed algorithm in simulated medical scenarios, we conducted data collection and experiment testing on the nasotracheal intubation robot system as shown in Fig. \ref{hardware}. We designed this system to autonomously insert a nasotracheal tube (NTT) through the nostril, thereby establishing a connection between the airway and oxygen concentrators. The control device of the system can manipulate the distal tip of the fiberoptic bronchoscope (FOB) in terms of its bending angle. The bronchoscopic view displays the field of vision from the camera on the distal tip part. The holistic feeding module is used to control the movement of the external parts of the FOB and the NTT.

The robotic arm remotely operates the control devices of the overall feeding module and FOB, gradually guiding the FOB tip from outside the phantom toward the nostril. The FOB enters the nasal cavity, passes through the pharynx, through the glottis opening, and into the trachea. Throughout this process, images of the bronchoscope view are continuously captured, including details such as the nostril, nasal passage, glottic slit, right glottic valve, and left glottic valve.

\subsection{Experimental Setup}

\begin{figure}[!t]
\centering
\subfloat[The typical samples of PID dataset]{
    \includegraphics[width=\columnwidth]{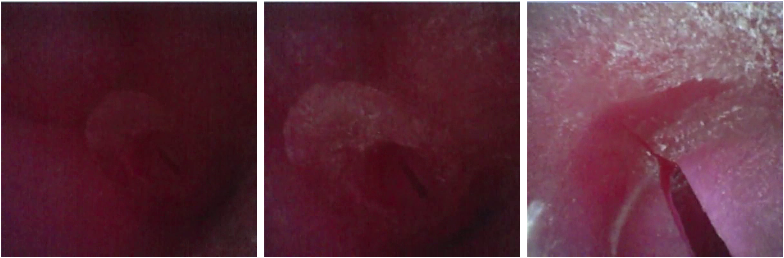}
}

\subfloat[The typical samples of BAGLS dataset]{
    \includegraphics[width=\columnwidth]{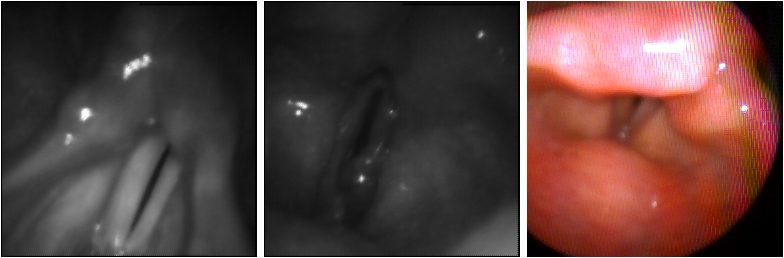}
}

\subfloat[The typical samples of clinical  dataset]{
    \includegraphics[width=\columnwidth]{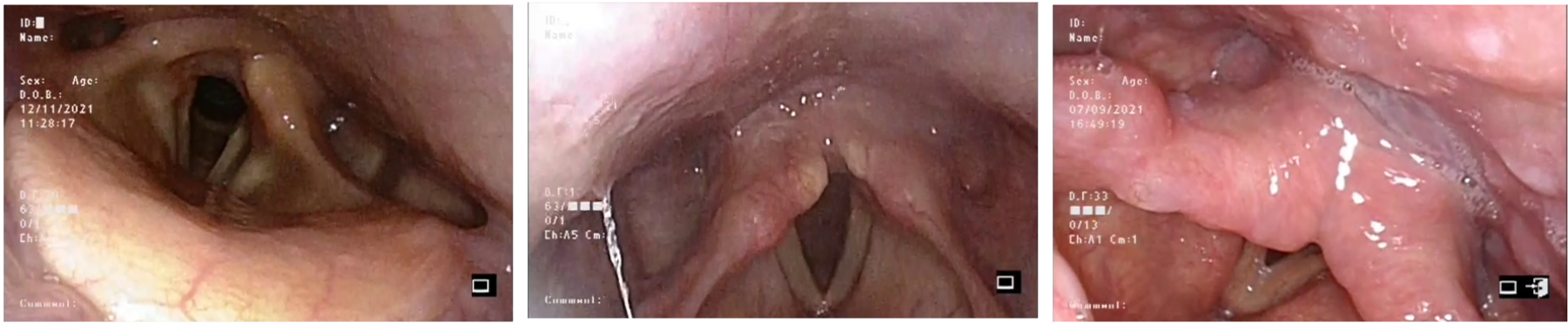}
}
\caption{The typical samples of datasets. Durin
g the NTI process, the glottis undergoes significant scale variations, and the environment around the glottis is complex with low light conditions. General detectors lose a lot of accuracy.}
\label{datasets}
\end{figure}

\subsubsection{Implementation Details}
We utilize the AdamW optimizer to train GlottisNet with the weight decay parameter set to 0.05.
For all ablation experiments, we set the batch size to 128 and conduct training for 500 epochs.
We implement a cosine annealing learning rate scheduler, initializing at 0.0005 and gradually attenuating to zero throughout the training progression.
To preserve the integrity of glottal structures during preprocessing, we implement dataset-specific scaling protocols: the phantom image dataset is resampled to $400\times 400$ pixels, the BAGLS dataset to $256\times 256$ pixels, and the clinical dataset to $640\times 640$ pixels.
In addition to conventional image preprocessing techniques such as scaling, cropping, padding, and flipping, we implement photometric distortion to enhance model robustness in low-light environments.

To ensure reproducibility and benchmark compatibility for all SOTA experiments across the three datasets, we strictly adhere to the hyperparameter configurations specified in~\cite{mmdet}, maintaining the original learning rate schedule, momentum coefficient, and regularization parameter without modification.
The experiments are performed on a computational platform equipped with an Intel Core i9-10980XE (3.0 GHz) processor, 128 GB system memory, and an NVIDIA RTX 3090 graphics processing unit (GPU).

\subsubsection{Datasets}
To systematically assess the efficacy and robustness of our method, we conducted comprehensive evaluations using three distinct datasets:

    \textbf{PID}: We developed the Phantom Image Dataset (PID) using phantoms to simulate diverse NTI scenarios as shown in Fig.~\ref{datasets}a. The dataset comprises 2,746 images with a resolution of $400\times 400$ pixels, partitioned into 2,267 training and 479 test samples. We meticulously annotated key anatomical structures, including the left nostril, bilateral glottal valves, and the glottic aperture, with bounding boxes and masks. These annotations effectively capture significant morphological variations.
    
 \textbf{BAGLS}: The BAGLS dataset~\cite{Dataset2} was preprocessed to extract bounding box annotations from the existing segmentation masks.  
    Subsequently, we converted the data to the conventional COCO~\cite{coco} format and divided it into training and test datasets containing 55,750 and 3,500 images, respectively.
    The BAGLS dataset contains images of 640 healthy and pathological people from high-speed video endoscopic recordings. 
    The recordings were acquired by multiple clinicians using diverse endoscopic equipment, thereby ensuring data heterogeneity.    
    
\textbf{Clinical}: The clinical dataset was acquired from standardized nasopharyngeal examination procedures conducted at Singapore General Hospital, ensuring consistency in data collection protocols~\cite{hao2024uaal}.
    These procedures are performed by board-certified otolaryngologists using commercial flexible nasopharyngeal endoscopes, thereby representing authentic clinical scenarios encountered during routine examinations. 
    The dataset comprises images extracted from high-definition examination videos of 82 adult patients, providing diverse anatomical representations.
    The anatomical structures annotated in the dataset for both detection and segmentation tasks include: bilateral nostrils, bilateral glottal valves, and glottal slit. 
    The dataset adheres to the COCO annotation format, with stratified partitioning into training and test sets containing 2728 and 1166 images, respectively, maintaining similar anatomical distribution across both subsets.
    This diverse dataset served as a crucial benchmark for validating the generalizability and clinical applicability of our model.

\subsubsection{Evaluation Metrics}
To comprehensively assess object detection performance, we employ standard COCO evaluation metrics~\cite{coco}, specifically the mean Average Precision (mAP) and mAP at an IoU threshold of 0.5 (AP50).
Furthermore, to quantitatively evaluate model robustness against scale variations, we incorporate scale-specific average precision metrics: $AP_S$ for small objects (area $< 32^2$ pixels), $AP_M$ for medium objects ($32^2 \le \text{area} < 96^2$ pixels), and $AP_L$ for large objects (area $\ge 96^2$ pixels).

For semantic segmentation tasks, we utilize metrics widely adopted in medical imaging, including the Intersection over Union (IoU) and the Dice coefficient. The multi-class segmentation performance is reported using the mean IoU (mIoU) and mean Dice (mDice). The mDice is calculated using the following equation:

\begin{equation}
    mDice = \frac{1}{n} \sum_{i=1}^n \frac{2 TP_i}{2 TP_i +FP_i + FN_i},
\end{equation}
where $TP_i$, $FP_i$, and $FN_i$ represent the true positives, false positives, and false negatives for the $i$-th class, respectively.
For an \textit{n}-class segmentation task, we calculate the Dice coefficient independently for each class and subsequently compute their arithmetic mean to obtain the final mDice score.
The computational efficiency of the model is quantitatively assessed through inference latency measurements, reported in FPS.
Additional computational metrics, including model size, parameter count, and floating-point operations per second (FLOPs), are critical for evaluating the deployability of the model on resource-constrained edge devices.
These performance indicators collectively form our comprehensive evaluation framework for model assessment.

\begin{table}[!t]
\centering
\renewcommand{\arraystretch}{1.1}
\caption{Ablation study for the overall network structure.}
\label{table1}
\begin{tabular}{rccccc}
\toprule
\multicolumn{1}{l}{}             & \multicolumn{2}{c}{Detection (\%)}               & \multicolumn{2}{c}{Segmentation(\%)}               & \multirow{2}{*}{\begin{tabular}[c]{@{}c@{}}model\\ size (MB)\end{tabular}} \\ \cmidrule(lr){2-3} \cmidrule(lr){4-5}
\multicolumn{1}{l}{}             & mAP                  & AP50              & mAP                  & mDice              &                                                                       \\ \midrule
\multicolumn{1}{l}{RTMDet-tiny} & 34.5         & 64.1          & --             & --            & 84.0          \\
*+ DWConv                        & 32.9         & 61.9          & --             & --            & \textbf{9.6}         \\
*+ LessShourCut                  & 34.0          & 61.1          & --             & --            & \textbf{9.6}         \\
*+ins-branch                      & 44.4         & 68.6          & 43.6          & 74.5         & 12.0          \\
*+DetHeads-free                  & 44.7         & 64.1          & 43.8          & 72.7         & 11.0          \\
*+LightSRM                & \textbf{46.0} & \textbf{75.8} & \textbf{44.2} & \textbf{81.1} & 19.0 \\ \bottomrule
\end{tabular}
\end{table}

\begin{table*}[!t]
    \centering
    \renewcommand{\arraystretch}{1.1}
    \caption{Ablation study for the LightSRM.}
    \label{LightSRM}
    \begin{tabular}{ccccccccc}
    \toprule
    \multirow{2}{*}{{\begin{tabular}[c]{@{}c@{}}Channel\\ Attention\end{tabular}}} & \multirow{2}{*}{DilatedConv} & \multirow{2}{*}{DWCM} & \multicolumn{2}{c}{Detection(\%)} & \multicolumn{4}{c}{Segmentation(\%)}                          \\ 
    \cmidrule(lr){4-5} \cmidrule(lr){6-9}  
                                                                                          &                              &                       & mAP             & AP50         & mAP           & AP50       & mIoU          & mDice         \\ \midrule
    \multirow{5}{*}{\ding{55}}                                                           & \ding{51}                    &                       & 32.4            & 52.0            & 32.1          & 51.4          & 41.6          & 58.8          \\
                                                                                          &                              & \ding{51}             & 32.5            & 50.9            & 32.4          & 51.8          & 43.6          & 60.7          \\
                                                                                          & \ding{51}                    & \ding{55}             & 33.9            & 53.9            & 34.0          & 57.0          & 47.9          & 64.8          \\
                                                                                          & \ding{55}                    & \ding{51}             & 37.4            & 61.0            & 36.6          & 60.0          & 48.8          & 65.6          \\
                                                                                          & \ding{51}                    & \ding{51}             & 39.6            & 62.3            & 40.1          & 63.9          & 53.8          & 70.0            \\ \hline
    \multirow{5}{*}{\ding{51}}                                                           & \ding{51}                    &                       & 39.8            & 60.2            & 38.9          & 61.0          & 51.5          & 68.0          \\
                                                                                          &                              & \ding{51}             & 40.5            & 61.1            & 39.8          & 61.5          & 52.9          & 69.2          \\
                                                                                          & \ding{51}                    & \ding{55}             & 39.8            & 60.7            & 38.8          & 60.7          & 50.5          & 67.1          \\
                                                                                          & \ding{55}                    & \ding{51}             & 40.1            & 61.1            & 38.8          & 60.7          & 52.4          & 68.8          \\
                                                                                          & \ding{51}                    & \ding{51}             & \textbf{46.0}   & \textbf{75.8}   & \textbf{44.2} & \textbf{75.0} & \textbf{68.2} & \textbf{81.1} \\ \bottomrule
    \end{tabular}
\end{table*}

\begin{table}[]
    \centering
    \renewcommand{\arraystretch}{1.1}
    \caption{Ablation study for the dilation rate.}
    \label{dilation}
    \begin{tabular}{ccccccc}
    \toprule
    \multirow{2}{*}{\begin{tabular}[c]{@{}c@{}}Dilation\\ rate\end{tabular}} & \multicolumn{2}{c}{Detection(\%)} & \multicolumn{4}{c}{Segmentation(\%)} \\ 
    \cmidrule(lr){2-3} \cmidrule(lr){4-7} 
                                                                             & mAP             & AP50            & mAP     & AP50    & mIoU   & mDice   \\ \midrule
    $[1]$                                                                 & 41.1            & 62.6            & 40.1    & 62.5    & 56.1   & 71.9    \\
    $[1, 2]$                                                               & 44.9            & 69.4            & 44.1    & 68.9    & 59.5   & 74.6    \\
    $[1, 2, 5]$                                                            & \textbf{46.0}   & \textbf{75.8}   & \textbf{44.2} & \textbf{75.0} & \textbf{68.2} & \textbf{81.1}    \\
    $[1, 2, 5 ,1 ,2]$                                                     & 43.7            & 68.2            & 42.9    & 64.6    & 53.7   & 69.9    \\ \bottomrule
    \end{tabular}
\end{table}

\begin{table}[!t]
    \centering
    \renewcommand{\arraystretch}{1.1}
    \caption{Ablation study for the cost matrix.}
    \label{table2}
    \begin{tabular}{@{}ccccccccc@{}}
    \toprule
    \multicolumn{3}{c}{Cost weights} & \multicolumn{2}{c}{Detection(\%)} & \multicolumn{4}{c}{Segmentation(\%)} \\
    \cmidrule(lr){1-3} \cmidrule(lr){4-5} \cmidrule(lr){6-9}
        $\lambda_{1}$                    & $\lambda_{2}$                 & $\lambda_{3}$                   & mAP         & AP50     & mAP         & AP50     & mIoU        & mDice       \\
        \midrule
        3                                & 1                             & 1                               & 54.0          & 79.6          & 52.2          & 81.0          & 80.0          & 88.9          \\
        1                                & 1                             & 3                               & 51.9          & 78.1          & 51.9          & 78.9          & 77.1          & 87.1          \\
        3                                & 1                             & 3                               & \textbf{55.2} & 81.0          & 53.9 & \textbf{81.6} & \textbf{81.5} & \textbf{89.8} \\
        3                                & 3                             & 1                               & 54.0          & 79.9          & 52.5          & 80.5          & 79.5          & 88.6          \\
        1                                & 3                             & 3                               & 52.1          & 79.2          & 51.2          & 78.7          & 76.7          & 86.8          \\
        1                                & 3                             & 1                               & 54.9          & \textbf{81.2} & \textbf{54.2} & 79.9          & 77.1          & 87.1          \\
        1                                & 1                             & 1                               & 41.8          & 62.6          & 41.0          & 61.2          & 53.4          & 69.6          \\
        3                                & 1                             & 0                               & 43.8          & 67.4          & 43.1          & 67.5          & 58.4          & 73.7          \\
        3                                & 0                             & 1                               & 41.1          & 66.5          & 41.5          & 66.8          & 58.5          & 73.8          \\
        0                                & 1                             & 3                               & 43.8          & 67.0          & 42.4          & 69.5          & 59.6          & 74.7          \\
        \bottomrule
    \end{tabular}
\end{table}

\begin{table}[!t]
    \centering
    \renewcommand{\arraystretch}{1.1}
    \caption{Ablation study for the TopK.}
    \label{Topk}
    \begin{tabular}{ccccccc}
    \toprule
    \multirow{2}{*}{TopK} & \multicolumn{2}{c}{Detection (\%)} & \multicolumn{4}{c}{Segmentation (\%)}                              \\ \cmidrule(lr){2-3} \cmidrule(lr){4-7}
    & mAP         & AP50     & mAP         & AP50     & mIoU        & mDice         \\ \midrule
    1                              & 52.2          & 80.5          & 52.5          & 82.2          & 80.0          & 88.9          \\
    3                              & \textbf{57.2} & \textbf{86.4} & \textbf{56.5} & \textbf{85.9} & \textbf{84.7} & \textbf{91.7} \\
    5                              & 55.8          & 81.0          & 54.8          & 83.3          & 82.3          & 90.3          \\
    7                              & 54.6          & 83.6          & 53.9          & 81.9          & 77.6          & 87.4          \\
    9                              & 54.8          & 81.4          & 53.7          & 81.3          & 79.4          & 88.5          \\
    11                             & 55.8          & 84.4          & 55.3          & 84.1          & 82.8          & 90.6          \\
    13                             & 55.2          & 81.0          & 53.9          & 81.6          & 81.5          & 89.8          \\
    15                             & 55.3          & 82.0          & 53.9          & 82.1          & 81.0          & 89.5          \\ \bottomrule
    \end{tabular}
\end{table}

\subsection{Ablation Studies}

\begin{table*}[!t]
    \centering
    \renewcommand{\arraystretch}{1.2}
    \caption{Comparison of accuracy with state-of-the-art methods on PID and BAGLS datasets.}
    \label{SOAT}
    \begin{tabular}{lcccccccccccc}
        \toprule \multirow{3}{*}{Methods}                                             & \multicolumn{6}{c}{PID dataset}    & \multicolumn{6}{c}{BAGLS dataset}      \\
        \cmidrule(lr){2-7} \cmidrule(lr){8-13}                                        & \multicolumn{2}{c}{Detection (\%)} & \multicolumn{4}{c}{Segmentation (\%)} & \multicolumn{2}{c}{Detection (\%)} & \multicolumn{4}{c}{Segmentation (\%)} \\
        \cmidrule(lr){2-3} \cmidrule(lr){4-7} \cmidrule(lr){8-9} \cmidrule(lr){10-13} & mAP                              & AP50                             & mAP                              & AP50                            & mIoU        & mDice       & mAP         & AP50     & mAP         & AP50     & mIoU        & mDice       \\
        \midrule ATSS~\cite{ATSS}                                                     & 30.3                               & 59.5                                  & --                                 & --                                   & --            & --            & 53.7          & 84.8          & --            & --            & --            & --            \\
        BoxInst~\cite{BoxInst}                                                        & 33.2                               & 53.4                                  & 13.4                               & 36.5                                 & 28.2          & 44.0          & 53.8          & 84.1          & 17.0          & 49.6          & 38.3          & 55.4          \\
        CondInst~\cite{CondInst}                                                      & 33.5                               & 49.7                                  & 30.0                               & 46.5                                 & 36.5          & 53.5          & 41.1          & 79.1          & 17.7          & 51.4          & 43.3          & 60.4          \\
        CMask RCNN~\cite{Cascade_RCNN}                                                & 40.2                               & 64.0                                  & 36.5                               & 62.0                                 & 48.7          & 65.5          & 55.9          & 83.2          & 46.7 & 82.8          & 72.0          & 83.7          \\
        Conditional DETR~\cite{Conditional_DETR}                                      & 15.6                               & 37.3                                  & --                                 & --                                   & --            & --            & 7.1           & 21.9          & --            & --            & --            & --            \\
        DAB DETR~\cite{DAB_DETR}                                                      & 18.4                               & 44.2                                  & --                                 & --                                   & --            & --            & 8.7           & 23.7          & --            & --            & --            & --            \\
        Deform. DETR~\cite{Deformable_DETR}                                           & 25.7                               & 52.0                                  & --                                 & --                                   & --            & --            & 38.4          & 74.7          & --            & --            & --            & --            \\
        DETR~\cite{DETR}                                                              & 6.7                                & 17.3                                  & --                                 & --                                   & --            & --            & 3.3           & 10.6          & --            & --            & --            & --            \\
        GFL~\cite{GFL}                                                                & 33.1                               & 61.3                                  & --                                 & --                                   & --            & --            & 54.2          & 83.8          & --            & --            & --            & --            \\
        Mask2Former~\cite{Mask2Former}                                                & 36.4                               & 51.9                                  & 33.3                               & 51.0                                 & 40.8          & 58.0          & 39.8          & 72.3          & 35.2          & 76.7          & 68.8          & 81.5          \\
        Mask RCNN (R50)~\cite{maskrcnn}                                               & 33.3                               & 54.9                                  & 33.5                               & 54.7                                 & 43.1          & 60.2          & 51.2          & 79.1          & 45.2          & 78.9          & 66.3          & 79.7          \\
        Mask RCNN (Swin-T)~\cite{maskrcnn}                                            & 27.4                               & 54.5                                  & 29.8                               & 53.1                                 & 43.3          & 60.4          & 53.8          & 79.4          & 46.0          & 79.1          & 66.4          & 79.8          \\
        MS RCNN~\cite{Mask-Scoring-RCNN}                                              & 36.7                               & 58.4                                  & 34.9                               & 57.2                                 & 44.7          & 61.8          & 48.2          & 75.9          & 45.1          & 77.3          & 64.3          & 78.3          \\
        PointRend~\cite{pointrend}                                                    & 32.3                               & 60.1                                  & 35.6                               & 63.4                                 & 53.6          & 69.8          & 50.2          & 78.4          & 46.5          & 79.8          & 67.8          & 80.8          \\
        QueryInst~\cite{queryinst}                                                    & 9.8                                & 19.8                                  & 11.0                               & 21.3                                 & 19.8          & 33.1          & 53.4          & 87.0          & 43.5          & 85.7 & 78.7 & 88.1 \\
        SOLOv2~\cite{SOLOv2}                                                          & --                                 & --                                    & 36.2                               & 57.3                                 & 45.5          & 62.5          & --            & --            & 23.1          & 56.3          & 40.1          & 57.2          \\
        SparseInst~\cite{SparseInst}                                                  & --                                 & --                                    & 30.8                               & 46.1                                 & 34.6          & 51.4          & --            & --            & 29.0          & 61.9          & 46.0          & 63.0          \\
        RTMDet\_s~\cite{lyu2022rtmdet}                                                & 40.6                               & 63.7                                  & 39.7                               & 61.9                                 & 56.9          & 72.5          & 55.4          & 87.5          & 19.6          & 55.1          & 42.6          & 59.7          \\
        U-Net\text{++}~\cite{unet++} & -- & -- & -- & -- & 53.0 & 69.3 & -- & -- & -- & -- & 78.3 & 87.8 \\
        U-Net\text{3+}~\cite{unet3+} & -- & -- & -- & -- & 37.0 & 54.0 & -- & -- & -- & -- & 77.3 & 87.2 \\
        YOLACT~\cite{YOLACT}                                                          & 40.0                               & 70.8                                  & 40.1                               & 68.9                                 & 58.2          & 73.6          & 38.5          & 71.7          & 27.2          & 60.8          & 48.5          & 65.3          \\
        YOLOv8n-seg~\cite{yolov8_u} & 46.7 & 77.5 & 18.2 & 42.5 & 31.1 & 47.4 & 49.6 & 75.5 & 22.7 & 58.7 & 42.2 & 59.3 \\
        YOLOv8s-seg~\cite{yolov8_u} & 49.8 & 77.4 & 19.4 & 46.4 & 34.0 & 50.8 & 51.7 & 76.3 & 23.7 & 61.1 & 43.8 & 60.9 \\
        YOLOv9c-seg~\cite{wang2024yolov9} & 41.5 & 69.2 & 18.1 & 40.1 & 30.5 & 46.8 & 51.0 & 74.0 & 22.4 & 58.7 & 44.8 & 61.8 \\
        YOLOv9e-seg~\cite{wang2024yolov9} & 44.8 & 69.6 & 20.3 & 48.4 & 33.4 & 50.1 & 53.0 & 73.8 & 21.8 & 56.0 & 43.2 & 60.3 \\
        YOLOv10n~\cite{yolov10_u} & 42.7 & 66.5 & -- & -- & -- & -- & 48.0 & 73.4 & -- & -- & -- & -- \\
        YOLOv10s~\cite{yolov10_u} & 43.5 & 71.5 & -- & -- & -- & -- & 50.9 & 75.4 & -- & -- & -- & -- \\
        YOLOv11n-seg~\cite{yolo11_u} & 43.5 & 74.0 & 19.4 & 49.1 & 35.9 & 52.8 & 50.5 & 76.2 & 23.2 & 59.7 & 42.1 & 59.3 \\
        YOLOv11s-seg~\cite{yolo11_u} & 49.3 & 79.4 & 20.0 & 47.8 & 33.4 & 50.1 & 51.2 & 77.4 & 23.2 & 60.7 & 42.8 & 59.9 \\
        \hline
        GlottisNet (Ours)                                                             & \textbf{57.8}                      & \textbf{87.2}                         & \textbf{56.7}                      & \textbf{87.1}                        & \textbf{86.7} & \textbf{92.9} & \textbf{60.8} & \textbf{91.8} & \textbf{47.9}          & \textbf{86.7}          & \textbf{81.2}          & \textbf{89.6}          \\
        \bottomrule
    \end{tabular}
\end{table*}

\begin{table*}[!t]
    \centering
    \begin{threeparttable}
    \renewcommand{\arraystretch}{1.2}
    \caption{Comparison of accuracy, model size, and FPS with state-of-the-art methods on the clinical dataset.}
    \label{SOAT_modelsize}
    \begin{tabular}{lcccccccccc}
    \toprule
    \multirow{3}{*}{Methods} & \multirow{3}{*}{Backbone} & \multicolumn{6}{c}{Clinical dataset} & \multirow{3}{*}{\begin{tabular}[c]{@{}c@{}}Model\\ size (MB)\end{tabular}} & \multicolumn{2}{c}{\multirow{2}{*}{\begin{tabular}[c]{@{}c@{}}FPS\\(f/s)\end{tabular}}} \\ 
    \cmidrule(lr){3-8}
    & & \multicolumn{2}{c}{Detection (\%)} & \multicolumn{4}{c}{Segmentation (\%)} & & & \\ 
    \cmidrule(lr){3-4} \cmidrule(lr){5-8} \cmidrule(lr){10-11}
    & & mAP & AP50 & mAP & AP50 & mIoU & mDice & & GPU & CPU \\ 
    \midrule
    ATSS~\cite{ATSS}                         & ResNet50                  & 17.8            & 39.5            & --            & --            & --            & --            & 251                                                                        & 52.9                   & 11.4                  \\
    BoxInst~\cite{BoxInst}                   & ResNet50                  & 15.3            & 38.9            & 4.7           & 16.7          & 12.8          & 22.7          & 287                                                                        & 46.8                   & 18.3                  \\
    CondInst~\cite{CondInst}                 & ResNet50                  & 36.7            & 64.3            & 28.1          & 60.5          & 50.3          & 66.9          & 578                                                                        & 46.5                   & 9.9                   \\
    CMask RCNN~\cite{Cascade_RCNN}           & ResNet50                  & 30.2            & 56.2            & 28.2          & 54.9          & 50.2          & 66.8          & 602                                                                        & 33.9                   & 1.0                   \\
    Conditional DETR~\cite{Conditional_DETR} & ResNet50                  & 10.5            & 34.1            & --            & --            & --            & --            & 512                                                                        & 40.4                   & 17.1                  \\
    DAB DETR~\cite{DAB_DETR}                 & ResNet50                  & 18.3            & 43.5            & --            & --            & --            & --            & 515                                                                        & 35.4                   & 15.5                  \\
    Deform. DETR~\cite{Deformable_DETR}      & ResNet50                  & 28.7            & 58.9            & --            & --            & --            & --            & 474                                                                        & 39.9                   & 3.5                   \\
    DETR~\cite{DETR}                         & ResNet50                  & 9.2             & 21.6            & --            & --            & --            & --            & 557                                                                        & 47.1                   & 20.0                  \\
    GFL~\cite{GFL}                           & ResNet101                 & 34.0            & 59.3            & --            & --            & --            & --            & 401                                                                        & 39.4                   & 8.6                   \\
    Mask2Former~\cite{Mask2Former}           & ResNet50                  & 32.6            & 59.8            & 28.0          & 56.8          & 47.7          & 64.6          & 539                                                                        & 21.0                   & 3.3                   \\
    Mask RCNN~\cite{maskrcnn}                & ResNet50                  & 19.8            & 44.4            & 23.8          & 46.3          & 41.6          & 58.8          & 344                                                                        & 51.9                   & 2.4                   \\
    Mask RCNN~\cite{maskrcnn}                & Swin-T                    & 20.9            & 47.4            & 22.7          & 47.2          & 43.7          & 60.8          & 556                                                                        & 40.2                   & 2.1                   \\
    MS RCNN~\cite{Mask-Scoring-RCNN}         & ResNet50                  & 19.1            & 46.3            & 24.6          & 48.4          & 44.0          & 61.1          & 471                                                                        & 48.0                   & 2.4                   \\
    PointRend~\cite{pointrend}               & ResNet50                  & 23.8            & 50.9            & 25.5          & 51.4          & 48.6          & 65.4          & 437                                                                        & 41.4                   & 2.4                   \\
    QueryInst~\cite{queryinst}               & ResNet50                  & 7.6             & 17.1            & 8.4           & 17.8          & 17.3          & 29.5          & 2021                                                                       & 29.8                   & --                    \\
    SOLOv2~\cite{SOLOv2}                     & ResNet50                  & --              & --              & 26.6          & 58.0          & 49.4          & 66.1          & 361                                                                        & 44.2                   & 4.9                   \\
    SparseInst~\cite{SparseInst}             & ResNet50                  & --              & --              & 20.1          & 42.0          & 31.1          & 47.4          & 410                                                                        & 73.4                   & 12.6                  \\
    RTMDet\_s~\cite{lyu2022rtmdet}            & CSPNeXt                   & 36.3            & 62.5            & 34.3          & 61.9          & 59.7          & 74.8          & 169                                                                        & 43.9                   & 22.1                  \\
    U-Net\text{++}~\cite{unet++}              & --                & --            & --            & --          & --          & 33.0          & 49.6          & 32                                                                        & 55.6                  & 2.6                  \\
    U-Net\text{3+}~\cite{unet3+}              & --                & --            & --            & --          & --          & 21.8          & 35.9          & 104                                                                        & 33.3                  & --                  \\
    YOLACT~\cite{YOLACT}                     & ResNet50                  & 16.8            & 42.6            & 18.0          & 39.3          & 35.9          & 52.8          & 274                                                                        & 74.9                   & 7.2                   \\
    YOLOv8n-seg~\cite{yolov8_u}              & CSPDarknet                & 22.1            & 49.6            & 17.2          & 45.1          & 32.1          & 48.6          & 6.5                                                                        & 149.3                  & 42.6                  \\
    YOLOv8s-seg~\cite{yolov8_u}              & CSPDarknet                & 25.1            & 52.8            & 20.3          & 50.1          & 36.3          & 53.2          & 23                                                                         & 138.9                  & 16.3                  \\
    YOLOv9c-seg~\cite{wang2024yolov9}        & GELAN                     & 28.1            & 56.4            & 21.6          & 50.6          & 35.4          & 52.3          & 54                                                                         & 153.8                   & 5.2                   \\
    YOLOv9e-seg~\cite{wang2024yolov9}        & GELAN                     & 24.3            & 53.1            & 19.4          & 48.5          & 33.0          & 49.6          & 117                                                                        & 81.3                    & 2.8  \\
    YOLOv10n~\cite{yolov10_u} & -- & 21.1 & 46.9 & -- & -- & -- & -- & \textbf{5.5} & 166.7 & 39.8 \\
    YOLOv10s~\cite{yolov10_u} & -- & 24.7 & 53.4 & -- & -- & -- & -- & 16 & 163.9 & 20.4 \\
    YOLOv11n-seg~\cite{yolo11_u}             & --                         & 22.6            & 51.3            & 17.1          & 46.6          & 33.2          & 49.8          & 5.7                                                                        & 156.3                   & 21.4                   \\
    YOLOv11s-seg~\cite{yolo11_u}             & --                        & 26.3            & 53.9            & 21.7          & 51.6          & 36.4          & 53.4          & 20                                                                         & 144.9                   & 17.6                  \\ \hline
    GlottisNet (Ours)                        & --                         & \textbf{37.2}   & \textbf{73.1}   & \textbf{35.2} & \textbf{70.9} & \textbf{63.4} & \textbf{77.6} & 19                                                                & \textbf{171.4}         & \textbf{44.4}         \\ \bottomrule
\end{tabular}
\begin{tablenotes}
\item[*]
For a fair comparison, we uniformly matched the image size to $400\times400$ when testing FPS.
This dimensional difference results in varying computational loads, affecting the inference speed across datasets. 
Additionally, variations in CUDA versions can impact the testing results. For this experiment, CUDA version 11.8 is utilized.
\end{tablenotes}
\end{threeparttable}
\end{table*}

\subsubsection{Overall Network Structure}
To systematically evaluate the proposed architecture, we conduct comprehensive ablation studies.
We employ RTMDet-tiny as our baseline model and evaluate its performance on the PID dataset (Table \ref{table1}).
The first row demonstrates that RTMDet-tiny achieves an mAP of 34.5\% with a model size of 84 MB. 
However, the baseline model suffers from two limitations: it lacks segmentation capabilities and incurs significant computational overhead due to its $3 \times3$ convolution-based dimensionality reduction before the detection head.
To reduce computational complexity, we implement depth-wise separable convolutions in place of standard convolutions.
This modification reduces the model size by a factor of 8.75 (to 9.6 MB), albeit with a slight decrease in accuracy.
To mitigate this performance drop, we optimize the network by reducing shortcut paths in the neck, enabling the detection head to better utilize semantic features. 
The optimized architecture maintains a compact 9.6 MB model size while achieving an mAP of 34.0\%.

To incorporate segmentation capabilities, we introduce a dedicated branch with minimal overhead, utilizing only two $1\times1$ convolutions for channel adjustment.
This architectural enhancement yields a substantial 130\% improvement in mAP performance with a minimal model size increase of 1.4 MB.
Concurrently, the model achieves a competitive mDice score of 72.7\%.
Furthermore, the newly designed LightSRM was introduced into the overall architecture. 
This structure is cascaded throughout both the backbone and neck to enhance feature extraction and fusion capabilities.
As shown in Table~\ref{table1}, the incorporation of LightSRM significantly boosts the $mAPs$ to 46.0\% and 44.2\%, respectively.
The LightCSP demonstrates its efficacy through a 1.1-fold improvement in both AP50 and mDice metrics, achieving 75.8\% and 81.1\%, respectively.

\subsubsection{LightSRM}
\label{LightSRM_sub_sub}
To rigorously evaluate the efficacy and novel contributions of the LightSRM architecture, we conduct comprehensive ablation studies.
As illustrated in Table~\ref{LightSRM}, the experimental design compares three distinct configurations to isolate the impact of channel attention integration, enabling systematic comparative analysis of model performance. 

We compared three configurations: (1) individual DilatedConv or DWCM with residual connections; (2) these modules without residuals to test path dependency; and (3) the proposed hybrid configuration combining both modules (final row).

Our quantitative analysis demonstrates that the ablation of channel attention results in a marked performance degradation across all experimental groups.
These findings substantiate the critical importance of preserving channel-specific feature representations in the final stages of the LightSRM computational pathway. 
The naive application of $1 \times 1$ convolution for dimensionality reduction introduces two detrimental effects: (1) it artificially amplifies linear inter-channel dependencies, and (2) it precipitates the attenuation of discriminative features in both the primary and residual pathways, ultimately culminating in substantial performance deterioration.

Furthermore, for the model variant lacking channel attention mechanisms, we observe that omitting the residual connection yields significantly higher performance compared to retaining it.
This finding offers compelling empirical validation for our proposed strategy of reducing shortcut connections, consistent with the results presented in Table~\ref{table1}.
After adding channel attention, the accuracy fluctuation caused by the shortcut path can be significantly reduced, making the model performance more stable.
Upon integration of channel attention mechanisms, the performance variability induced by shortcut connections is substantially mitigated, resulting in enhanced model stability across evaluation metrics.

Finally, the last row of Table~\ref{LightSRM} demonstrates that the simultaneous integration of DilatedConv, DWCM, and the attention mechanism yields optimal network performance, substantiating the efficacy of our LightSRM design.

\subsubsection{Dilation rate}
\label{Dilation}
Considering that dilated convolutions inherently risk of information loss between sampled points, we implement a cascade of multiple dilated convolution layers with systematically varying dilation rates to maintain an extensive receptive field while substantially reducing computational complexity.
Following the dilation rate configuration strategy proposed by~\cite{Dliate}, we conduct comprehensive ablation studies examining various dilation rate combinations to quantitatively evaluate their performance impact.

As demonstrated in Table~\ref{dilation}, when the dilation rates of the cascaded dilated convolutions fall below the configuration $[1, 2, 5]$, a significant degradation in detection accuracy occurs.
This performance decline is attributable to insufficient receptive field coverage during the sparse sampling process, which results in substantial information loss in the inter-sample regions.
Conversely, employing excessive cascaded dilated convolutions results in oversampling of the input space.
This configuration not only introduces computational redundancy but also adversely impacts detection precision, contradicting the fundamental efficiency objectives of our proposed LightSRM.

\subsubsection{Cost Matrix}
\label{Cost_m}

\begin{figure}[!t]
\centerline{\includegraphics[width=\columnwidth]{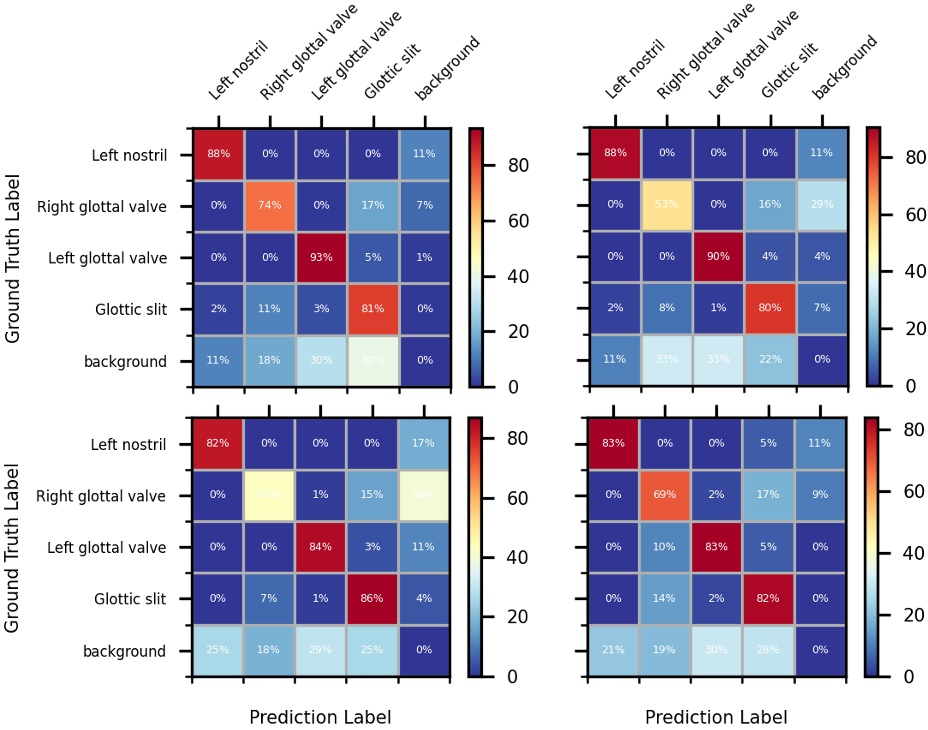}}
\caption{Confusion matrix of the models with different cost weights: $[3, 1, 3]$, $[3,1,0]$, $[3,0,1]$, and $[0,1,3]$. 
The omission of any single cost component results in missed detection of the hard-to-classify samples, such as the right glottal valve. This highlights that our designed cost matrix $[3, 1, 3]$  effectively reduces intra-class variation among these challenging samples. }
\label{CM}
\end{figure}
We evaluate the effectiveness of the cost matrix by analyzing its impact on model performance metrics.
Using our best-performing model configuration (Table~\ref{table1}, last row), we conduct comprehensive ablation studies on the PID dataset (Table~\ref{table2}).
The initial experiment optimizes the classification cost while maintaining the cost matrix.
This optimization results in significant improvements, yielding a detection mAP of 54.0\% and a mDice of 88.9\%.
Progressive weight adjustments of individual cost components, maintaining a fixed ratio of $3:1$, result in an mAP of 55.2\% and a mDice of 89.8\% when the cost matrix is configured as $[3, 1, 3]$.
This result represents a notable improvement over the pre-optimization accuracy, demonstrating the effectiveness of this weighting strategy in enhancing detection performance.
The last three rows of Table~\ref{table2} demonstrate that omitting any cost component results in significant performance degradation, validating the necessity of our cost matrix design.

Furthermore, comparative analysis between the first three rows and the middle three rows demonstrates that increasing the classification cost weight in the cost matrix leads to improved detection accuracy at the expense of reduced segmentation precision.
This result directly correlates with the architectural design of detection process of the GlottisNet.
Specifically, GlottisNet employs a sequential approach: first performing classification and box regression, followed by mask prediction within the defined bounding box.
Consequently, during mask prediction, the model can de-emphasize classification cost and prioritize spatial localization accuracy.

In our designed cost matrix, the $C_{cls}$ enhances inter-class differentiation, effectively reducing classification complexity and subsequently improving bounding box prediction accuracy. 
The $C_{IoU}$ and $C_{cen\_pre}$ mitigate intra-class variations, enabling the model to effectively differentiate between background and foreground elements in complex NTI scenarios while preserving sharp boundary delineation during segmentation.
To optimize segmentation performance while maintaining balanced detection capabilities, we empirically determined that a $[3, 1, 3]$ weight configuration in the cost matrix yields optimal results.
The ablation study presented in the last three rows of Table~\ref{table2} conclusively demonstrates that the omission of any cost component results in substantial performance degradation, thus validating the necessity and efficacy of our comprehensive cost matrix design.

To visualize the effectiveness of our proposed cost function in reducing intra-class variation, we analyze confusion matrices from models with different cost weight configurations: the baseline $[3, 1, 3]$ and last three ablated versions from Table~\ref{table2}, shown in Fig.~\ref{CM}.
The right glottal valve and glottic slit represent particularly hard-to-classify samples due to their spatial proximity and visual similarity at small scales.
The ablation of any cost function component significantly degrades the ability of the model to classify these anatomical structures, thereby increasing both misclassification and detection failures.
Our complete cost matrix with optimized weights substantially reduces misclassification errors and improves detection accuracy for these challenging cases, effectively minimizing intra-class variation while maximizing inter-class separability.

\subsubsection{Number of Samples}
\begin{figure}[!t]
\centerline{\includegraphics[width=3.4in]{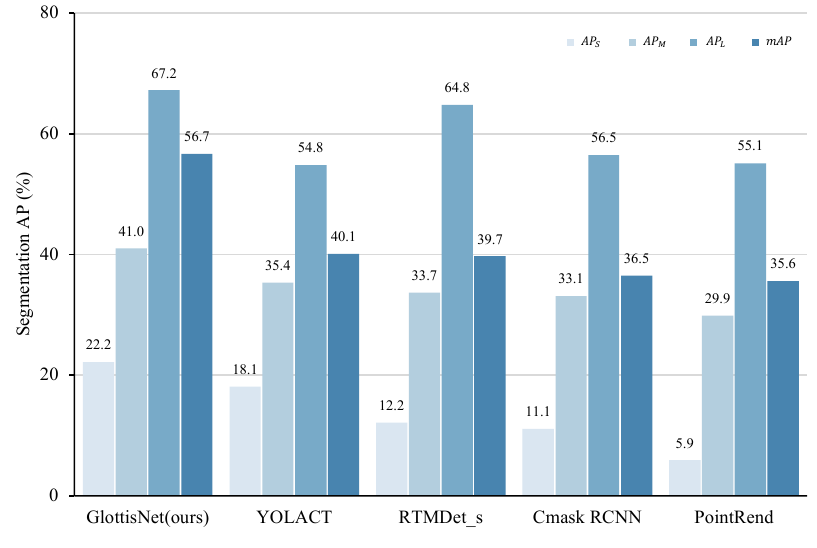}}
\caption{Quantitative scale robustness analysis on the PID dataset comparing detection accuracy across object scales against representative SOTA methods. GlottisNet demonstrates a significant advantage within the small-object category $AP_S$ by outperforming RTMDet-s with 22.2\% versus 12.2\%. These results validate the efficacy of the LightSRM module in handling drastic scale variations. }
\label{scale_robustness}
\end{figure}

Our label assignment method dynamically matches positive and negative samples within a predefined number of total samples.
Therefore, we conduct an ablation study on the required number of positive samples (TopK) by utilizing a cost matrix optimized with weights $[3,1,3]$ and initialized with 13 positive samples.
As shown in Table~\ref{Topk}, introducing additional positive samples beyond the baseline leads to a slight decrease in mDice performance to 89.5\%.
In contrast, strategically reducing the TopK yields substantial improvements in mDice performance metrics.
This finding indicates that the current positive sample set contains an excessive number of low-quality negative samples. 
Reducing the positive sample count to three results in significant performance gains, achieving an mAP of 57.2\% and mDice of 91.7\%.
A further reduction to a single positive sample leads to substantial performance deterioration, which is attributed to insufficient positive samples compromising the generalization capacity of the model.
Ultimately, we establish the optimal TopK as three in our final model configuration.

\begin{figure}[!t]
\centerline{\includegraphics[width=3.4in]{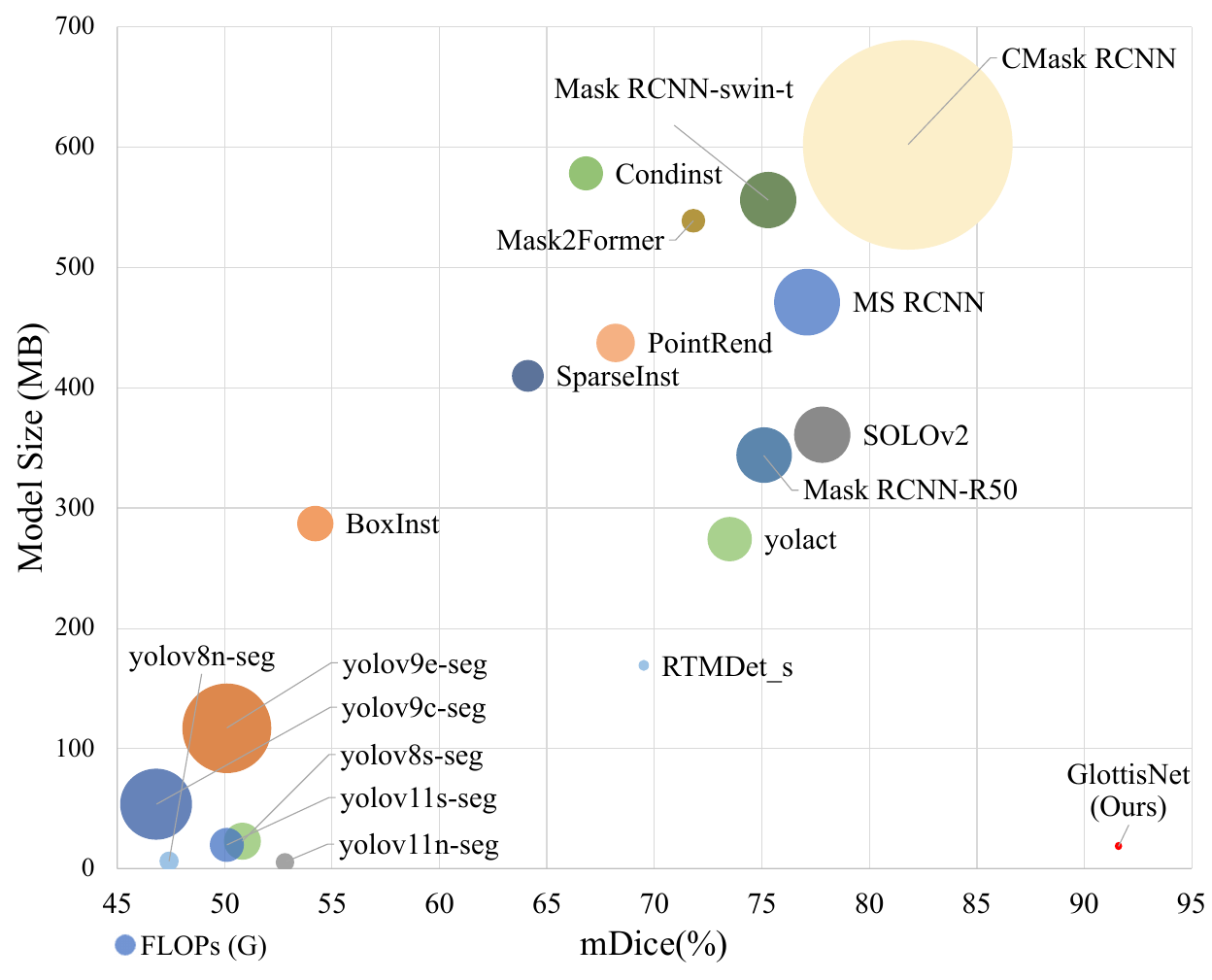}}
\caption{Comparing with state-of-the-art methods in terms of FLOPs, mDice, and model size. Our model achieves the highest accuracy while having the lowest computational cost.}
\label{Bubble}
\end{figure}

\begin{figure*}[!t]
\centerline{\includegraphics[width=6.9in]{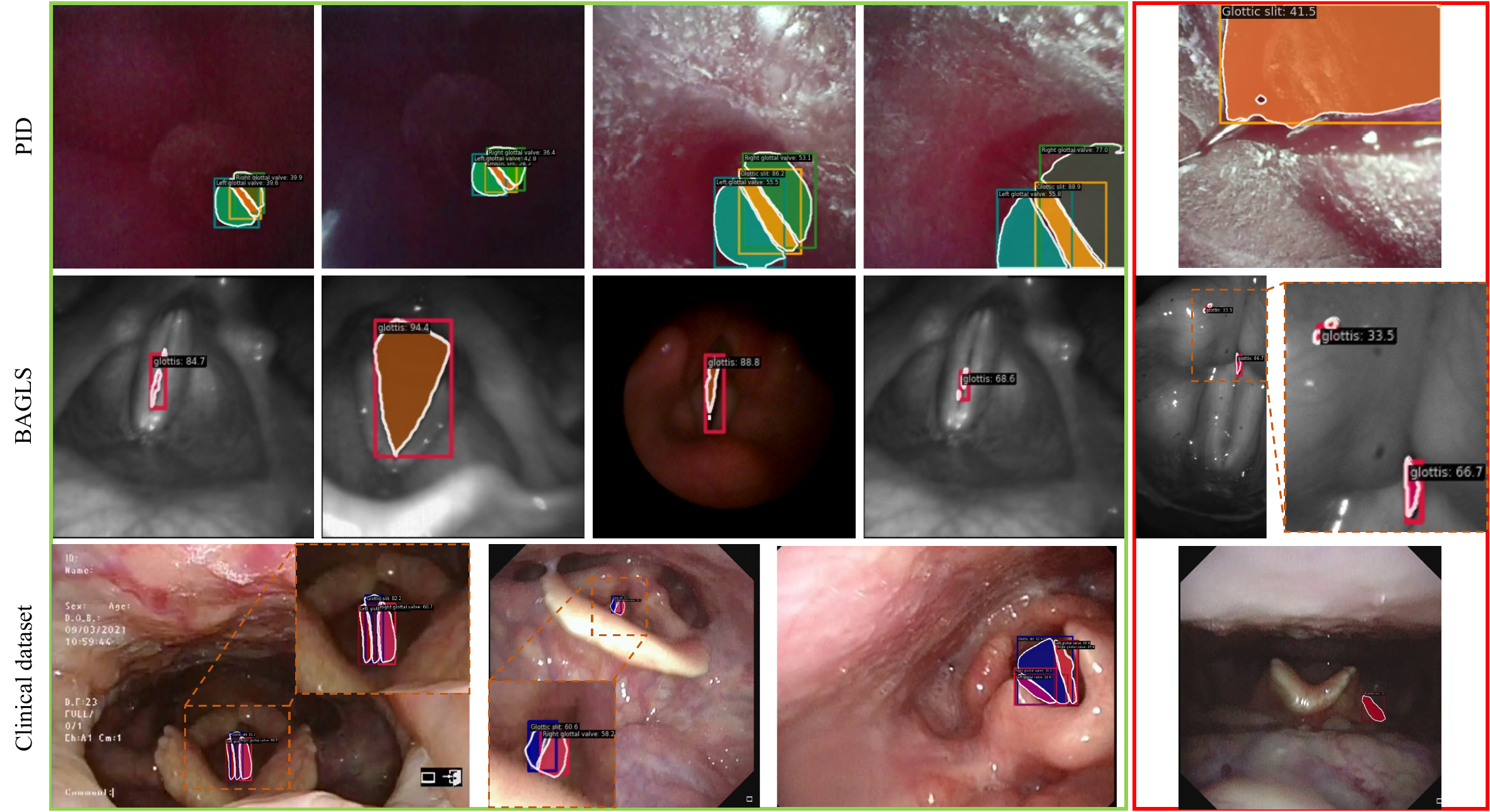}}
\caption{Qualitative results of our method are presented. 
The first few columns display correct detection results, while the last column illustrates incorrect detections.
Our method occasionally produces false positives under challenging conditions, particularly when the glottis is fully occluded or not visible, and when the imaging environment presents complex artifacts or noise. 
In future work, we will propose new methods to further solve this problem.}
\label{res}
\end{figure*}

\subsection{Comparison with State-of-the-Art Methods}
To quantitatively evaluate the performance of our proposed method, we conduct a comprehensive comparison with SOTA methods using the three datasets mentioned above.

\subsubsection{Accuracy}
As illustrated in Table~\ref{SOAT}, GlottisNet demonstrates superior detection performance with photometric distortion preprocessing, achieving mAP scores of 57.8\%, 63.1\%, and 37.2\% on these datasets, respectively, thereby surpassing all current SOTA methods.
When evaluated using the AP50 metric, all SOTA methods exhibit satisfactory performance on the structurally simpler BAGLS dataset.
However, performance declines substantially when algorithms are evaluated on the more challenging PID dataset and the high-complexity clinical dataset, as reflected by diminished AP50 scores across all methods.
Despite this general performance degradation, GlottisNet maintains an AP50 score exceeding 80\% on the PID dataset, outperforming the second-best method by a margin of 7\%.
Similarly, GlottisNet achieves superior performance on the clinical dataset, despite the significant challenges posed by environmental variability and anatomical diversity in these images.
These results substantiate the robust feature extraction capabilities of GlottisNet and superior generalization across diverse datasets, enabling more precise bounding box localization to facilitate navigation during NTI procedures across varying clinical scenarios.

In terms of segmentation performance, the mAP scores of various SOTA methods reveal significant limitations when applied to the complex PID and clinical datasets.
Specifically, only YOLACT achieves an mAP exceeding 40\% on the PID dataset, while only RTMDet-s attains an mAP above 30\% on the challenging clinical dataset.
These findings indicate that detector accuracy is substantially impacted by both the complexity of surgical environments and variations in glottic morphology across different clinical scenarios.
In contrast, GlottisNet achieves superior mAP scores of 56.7\% and 35.2\% on the PID and clinical datasets, respectively, significantly outperforming all compared SOTA methods.
A parallel performance pattern is observed with the mDice metric.
GlottisNet surpasses the second-best method by margins of 18.1\% and 2.8\% on the PID and clinical datasets, respectively.
These quantitative advantages demonstrate the superior feature extraction capabilities of GlottisNet in complex clinical environments with variable imaging conditions.

On the BAGLS dataset, which presents greater variations in glottic scale, only a limited number of SOTA methods, such as the Mask R-CNN family and PointRend, achieve mAP scores exceeding 40\%.
Similarly, for the mDice metric on this dataset, only certain two-stage detection algorithms demonstrate satisfactory segmentation accuracy.
In contrast, lightweight single-stage architectures, including the YOLO family, exhibit suboptimal performance across these metrics.
GlottisNet, however, achieves SOTA performance with mAP and mDice scores of 47.9\% and 89.6\%, respectively, surpassing all benchmark methods.
These results validate the enhanced scale invariance of GlottisNet compared to conventional lightweight architectures, demonstrating superior adaptability to glottic scale variations encountered during NTI procedures.

\subsubsection{Model size and FPS}
The quantitative analysis presented in Table~\ref{SOAT_modelsize} indicates that GlottisNet achieves significant parameter efficiency with a compact size of 19 MB, representing an 8-fold reduction compared to the baseline architecture.
With a fixed input resolution of $400 \times400$ pixels, the GlottisNet model demonstrates excellent inference performance on a single NVIDIA RTX 3090 GPU, achieving an overall processing throughput (including pre-processing and post-processing steps) of 171.4 FPS with a batch size of 1, which substantially exceeds the requirements for real-time detection applications.

It is noteworthy that while conventional clinical endoscopic systems typically operate at acquisition rates of 30-60 FPS, our model architecture is deliberately engineered for computational efficiency. This design choice preemptively addresses the speed degradation that inherently occurs when transitioning from high-performance GPUs to resource-constrained edge computing platforms in clinical settings.
This strategic optimization enables our system to maintain clinically viable performance on target hardware, even though the required processing capabilities considerably exceed the native acquisition frequency of current endoscopic imaging systems.
As illustrated in Table~\ref{SOAT_modelsize}, among all the SOTA models evaluated, only YOLOv8n-seg and YOLOv10n achieve detection speeds exceeding 30 FPS when deployed on CPU hardware.
The majority of models experience a significant performance degradation of over 80\% in processing speed when executed on CPU compared to their GPU counterparts, highlighting substantial computational constraints.

\subsubsection{Robustness Analysis}
To quantitatively substantiate the robustness of the proposed framework against scale variations visualized in Fig.~\ref{datasets} and to validate its efficacy in NTI scenarios, we conducted a stratified performance analysis employing scale-specific COCO metrics on the PID dataset.
We selected this dataset for the analysis because it simulates the complete visual trajectory of the intubation procedure and captures the continuum of glottal scales ranging from the initial distant entry to the final close-range insertion.

As illustrated in Fig.~\ref{scale_robustness}, GlottisNet exhibits consistent superiority across all scales. In the challenging small-object category $AP_S$ where the glottis appears as a minute target during the critical early phase of navigation, GlottisNet achieves a detection accuracy of 22.2\%. 
This performance surpasses the widely used real-time detector RTMDet-s at 12.2\% and the two-stage baseline Cascade Mask R-CNN at 11.1\% by a substantial margin.
Comparative analysis with YOLACT which demonstrates relatively better scale adaptability reveals that GlottisNet maintains a performance lead of 4.1\%. 
This near two-fold improvement in small-object detection confirms that the multi-receptive field design of the LightSRM module effectively mitigates feature loss for distant anatomical targets.

Furthermore, the superior performance of GlottisNet on the low-light Clinical dataset presented in Table~\ref{SOAT_modelsize} complements these findings by providing a systematic evaluation of illumination robustness.
The model outperforms SOTA methods by over 16\% in mAP on the clinical benchmark.
This empirical evidence confirms the resilience of the proposed framework to both geometric and photometric variations in real NTI scenarios.

\section{Discussion}
\label{Discussion}
This work presents GlottisNet, a real-time, high-precision framework designed to enhance safety and efficiency in NTI. This demanding procedure requires accurate and rapid glottis localization to prevent patient trauma, prolonged duration, and hypoxemia. 
In this context, although visual foundation models such as SAM and MedSAM have shown strong potential in general-purpose vision tasks, their substantial parameter scale and high inference latency limit practical deployment in real-time, resource-constrained clinical environments. 
GlottisNet serves as a robust visual navigation aid, especially in such settings, by overcoming challenges such as complex anatomy, glottal scale variations, and the accuracy-speed trade-offs of existing methods.

To accomplish this objective, GlottisNet implements a novel, computationally efficient architecture.
Unlike traditional segmentation networks such as U-Net that suffer from high latency, or general object detectors like the YOLO series that often sacrifice segmentation details for speed, GlottisNet employs a coupled detection-then-segmentation methodology. This design allows it to surpass the accuracy of heavy models such as Mask R-CNN while maintaining inference speeds exceeding 170 FPS. Such a performance metric is difficult for existing dual-stage architectures to match on standard hardware.
The framework initially conducts rapid localization of bounding boxes that encompass the glottal region, subsequently performing high-precision segmentation exclusively within these identified regions of interest. This architectural design transcends mere technical optimization, representing the cornerstone of practical clinical implementation. It enables GlottisNet to achieve real-time performance on standard CPU platforms, facilitating seamless integration into existing endoscopic equipment without necessitating expensive, specialized GPU hardware. This substantially reduces adoption barriers across diverse healthcare institutions, including hospitals and clinics with limited computational resources.

The substantial clinical contribution of GlottisNet lies in its inherent redundancy mechanisms, which significantly enhance procedural safety. The framework simultaneously delivers precise segmentation masks and robust detection bounding boxes, providing dual-modal output. In challenging clinical scenarios, including transient glottal closure or obstruction by secretions, segmentation masks can fail. During these critical instances, the detection head continues to provide a stable bounding box, ensuring uninterrupted navigational guidance for clinicians or robotic intubation systems. This redundant capability is essential for preserving clinician confidence while ensuring continuous, dependable guidance throughout the most technically challenging phases of intubation. As demonstrated in our experimental validation (Table~\ref{SOAT} and \ref{SOAT_modelsize}), existing dual-task architectures fail to deliver comparable reliability in real-time CPU environments, conferring a decisive advantage upon GlottisNet for clinical deployment.

Our comprehensive experimental evaluation validates the exceptional efficiency-performance balance achieved by GlottisNet. This superiority is directly attributable to the LightSRM module which captures fine-grained anatomical features across drastic scale changes. Such a capability is often lacking in generic lightweight backbones.
As demonstrated in Fig.~\ref{Bubble}, the proposed model achieves SOTA performance on segmentation tasks while requiring minimal computational resources, characterized by the lowest FLOPs and most compact model size among compared methods. 
This efficiency is paramount for resource-constrained mobile NTI systems, such as bedside equipment and training simulators. Furthermore, it enables deployment on portable devices for emergency interventions like wilderness rescue operations.

Despite its robust performance across diverse datasets (Fig.~\ref{res}), we acknowledge the limitations of GlottisNet. Under conditions of severe anatomical occlusion or obstruction, the model may occasionally generate false positive or false negative predictions. From a clinical safety perspective, false positive predictions could potentially misdirect surgical instrumentation, emphasizing the critical importance of deploying this system as a clinical decision-support tool to augment, rather than supersede, the expert judgment of experienced clinicians. Future research directions will concentrate on enhancing model robustness against these challenging clinical scenarios, with particular emphasis on occlusion-resistant algorithms. Our long-term objective encompasses the deployment of GlottisNet on edge computing platforms (e.g., NVIDIA Jetson Nano), establishing the foundation for next-generation intelligent, portable, and universally accessible NTI navigation systems. 
Building upon this visual foundation, our subsequent research will expand to system-level integration metrics, specifically addressing hand-eye calibration and closed-loop control stability. By bridging the gap between algorithmic perception and robotic execution, we aspire to deliver a safe, autonomous clinical tool that enhances patient outcomes and procedural success rates in NTI interventions.

\section{Conclusion}
\label{conclusion}
This paper proposes GlottisNet, a lightweight and real-time model designed to resolve the critical trade-off between segmentation precision and computational efficiency in robot-assisted NTI. Our findings demonstrate that the proposed architecture effectively mitigates the impact of drastic scale variations and complex anatomical environments without relying on computationally intensive models. By achieving SOTA accuracy with a model size of only 19 MB and inference speeds exceeding 170 FPS on a single GPU, GlottisNet proves that precise visual navigation is feasible for real-time robotic applications. Clinically, this implies that automated NTI assistance can be extended to resource-limited scenarios, potentially reducing the incidence of severe complications such as hypoxemia and airway trauma. Future work will focus on optimizing the framework for embedded hardware to support portable deployments and validating its robustness in broader clinical trials involving diverse pathological conditions.

\section{Acknowledgement}
The authors would like to express their gratitude to Dr. C. P. L. Chan and Dr. Jason Y. K. Chan from the Department of Otorhinolaryngology, Head and Neck Surgery, The Chinese University of Hong Kong, Hong Kong SAR, China, for their clinical insights and validation of the medical definitions used in this study.

\bibliographystyle{IEEEtran}
\bibliography{ref}

\end{document}